\documentclass[showpacs,preprint,amsmath,PRB]{revtex4}
\usepackage{graphicx}
\usepackage{dcolumn}
\usepackage{bm}

\begin{document}

\title{Probing Landau quantisation with the presence of insulator-quantum Hall transition in a GaAs two-dimensional electron system}

\author{Kuang Yao Chen, Y H Chang and C-T Liang$^{\ast}$}

\affiliation{Department of Physics, National Taiwan University, Taipei 106, Taiwan, R.O.C.}

\author{N Aoki and Y Ochiai$^{\ast}$}

\affiliation{Department of Electronics and Mechanical Engineering, Chiba University, Chiba 263, Japan}

\author{C F Huang}

\affiliation{National Measurement Laboratory, Centre for Measurement Standards, Industrial Technology Research Institute, 
Hsinchu 300, Taiwan, R.O.C.}

\author{Li-Hung Lin}

\affiliation{Department of Applied Physics, National Chiayi University, Chiayi 600, Taiwan, R.O.C.}

\author{K A Cheng}

\affiliation{Department of Electronic Engineering, Lung-Hwa University of Science and Technology, Taoyuan 333, Taiwan, R.O.C.}

\author{H H Cheng}

\affiliation{Centre for Condensed Matter Sciences, National Taiwan University, Taipei 106, Taiwan, R.O.C.}

\author{H H Lin}

\affiliation{Department of Electrical Engineering, National Taiwan University, Taipei 106, Taiwan, R.O.C.}

\author{Jau-Yang Wu and Sheng-Di Lin}

\affiliation{Department of Electronics Engineering, National Chiao Tung University, Hsinchu 300, Taiwan, R.O.C.}

\date{\today}

\begin{abstract}
Magneto-transport measurements are performed on the two-dimensional electron system (2DES) in an AlGaAs/GaAs 
heterostructure. By increasing the magnetic field perpendicular to the 2DES, magnetoresistivity oscillations 
due to Landau quantisation 
can be identified just near the direct insulator-quantum Hall (I-QH) transition. However, different mobilities 
are obtained from the oscillations and transition point. Our study shows that the direct I-QH transition does not 
always correspond to the onset of strong localisation. \newline
$^{\ast}$ ctliang@phys.ntu.edu.tw and ochiai@faculty.chiba-u.jp
\end{abstract}

\maketitle

The insulator to quantum Hall (I-QH) transition in a two-dimensional electron system (2DES) at low perpendicular magnetic fields $B$ has attracted much 
attention \cite{klz, song, lee, hanein, huckestein, liu, sheng, cfhuang, kim, tyhuang, nita1}. 
Theoretically, the direct I-QH transition from the insulator to an integer QH state of  $\nu \neq 1 $ is forbidden in an infinite, non-interacting 2DES with arbitrary amount of disorder, where $\nu$ is the Landau level filling 
factor \cite{klz, song, lee}. In such a system, the only allowed state at $B=0$ is the insulating one, and the 2DES 
undergoes the I-QH transition to enter the $\nu =1$ QH state \cite{jiang, hughes}. Realistically, however, only systems of finite sizes are available, and the effects of the electron-electron (e-e) interaction are significant in some 2DESs \cite{hanein, huckestein, li, cho, minkov, minkov1, arapov}.  As a result, the 2DESs may experience the direct I-QH transition from the low-field insulator to QH states of higher filling factors \cite{song, lee, hanein, cfhuang, minkov, minkov1, arapov}. Such a transition can be related to the zero-field metal-insulator transition, to which e-e interaction cannot be ignored \cite{hanein}. Given that most 2DESs show metallic behavior at $B$=0, the investigation of the direct I-QH transition at low $B$ should be conducted in low-mobility 2DESs \cite{klz, jiang}. 

The mechanisms for the direct I-QH transition are still under debate \cite{huckestein, liu, sheng, nita1, minkov1, arapov}. Huckestein \cite{huckestein} argued that such a transition is a crossover from weak localisation to Landau quantisation rather than a phase transition. Therefore the observed transition or crossing point is not a critical point.
According to Huckestein's argument, such a point should occur as the product 

\begin{eqnarray}
\mu B = 1.
\end{eqnarray}
Here $\mu$ is the mobility such that the strong localisation due to high-field Landau quantisation becomes important when the product $\mu B$, which equals the ratio of Landau-level spacing to broadening, is large enough. To be a measure for Landau quantisation, $\mu$ should be the quantum mobility. Because the strong localisation is believed to be important to the QH liquid, it seems natural that a 2DES undergoes the direct I-QH transition at $\mu B=1$ as we increase the perpendicular magnetic field. However, experimental evidence of quantum phase transition has been observed near the transition point \cite{cfhuang}. In addition, the existence of Landau quantisation in the low-field insulator indicates that its onset may be irrelevant to such a transition \cite{kim, tyhuang}. In fact, Landau quantisation could be unimportant to the crossover because its feature is absent near the crossing point in some reports \cite{li, cho}. Corrections based on the e-e interaction \cite{li, cho, minkov, arapov, kaminskii} and low-field Landau quantisation effects \cite{kim, tyhuang, nita1} are discussed in the literature. On the other hand, magneto-oscillations due to Landau quantisation appear just near the direct I-QH transition with increasing $B$ in some reports \cite{song, lee, cfhuang}. Huckestein's argument seems correct if we identify the onset of Landau quantisation by the appearance of magneto-oscillations. The observations of 
\begin{eqnarray}
\rho_{xy}/ \rho_{xx} \approx 1,
\end{eqnarray}
near the transition points \cite{song, lee} are also consistent with Huckestein's argument because $\rho_{xy}/\rho_{xx} = \mu B$ in the Drude model if the transport and quantum mobilities are the same. Here $\rho_{xx}$ and $\rho_{xy}$ are the longitudinal and Hall resistivities, respectively. To understand the direct I-QH transition, therefore, we shall re-examine the 2DESs where Landau quantisation induces oscillations just near the transition point occurring as Eq. (2) becomes valid with increasing $B$.  
In this study, we report a magneto-transport investigation on the 2DES in an AlGaAs/GaAs heterostructure. With increasing magnetic field $B$, amplitudes of resistivity oscillations $\Delta \rho_{xx}$ following the Shubnikov-de Haas (SdH) 
formula \cite{coleridge, martin, adamov, hang, jhchen}

\begin{eqnarray}
\Delta \rho_{xx} \propto \frac{\chi}{\mathrm{sinh}\chi}\mathrm{exp}(\frac{-\pi}{\mu B})
\end{eqnarray}
with $\chi=4\pi^{3}km^{\ast}T/heB$ can be identified just as the 2DES undergoes the direct I-QH transition. Here $T$ is the temperature, $k$, $h$, $e$, and $m^{\ast}$ are denoted as Boltzmann constant, Plank constant, electron charge, and effective mass, respectively. The oscillations are features of Landau quantisation, so it seems that the observed direct transition occurs near the onset of Landau quantisation just as suggested by Huckestein. In addition, Eq. (2) is valid at the transition point. However, different mobilities should be introduced just as in Refs.\cite{li, cho} 
because $\mu B$ is much smaller than 1 at the crossing point. One is for the direct I-QH 
transition and the other is for Landau quantisation. Therefore, corrections to Huckestein's argument should be taken into account even when the onset of Landau quantisation can be approximated by the transition point where Eq. (2) is valid.

The sample used in this study is an AlGaAs/GaAs heterostructure. Figure 1 shows its structure, where some Si atoms are doped in the 20-nm-wide GaAs quantum well to serve as the scattering sources. It is known that we can suppress the mobility to probe the integer quantum Hall effect by deliberately introducing some scattering sources in the quantum wells \cite{lee, kim, tyhuang}. The sample is made into the Hall pattern with the channel width 80 $\mu m$ by standard optical lithography, and AuGeNi alloy is annealed at 450 $^{0}$ C to fabricate the ohmic contacts. The magneto-transport measurements are performed in a top-loading He$^{3}$ system with the superconducting magnet.

Figure 2 shows the curves of the longitudinal resistivity $\rho_{xx}(B)$ at different temperatures and Hall resistivity $\rho_{xy}(B)$ at the temperature $T=4$~K under a low-frequency AC driving current of 40 nA. At low $B$, the 2DES behaves as an insulator such that $\rho_{xx}$ increases with decreasing $T$. The insulator is terminated at $B=3.5$~T $\equiv B_{c}$,  and $\rho_{xx}$ decreases with decreasing $T$ at $B > B_{c}$. Therefore, $B_{c}$ is the transition point. The filling factor $\nu \sim 8$ at $B_{c}$, and oscillations periodic in $1/B$ are observed when the sample behaves as a QH liquid at $B>B_{c}$. From the oscillating period in $1/B$, the carrier concentration $n=6.8 \times 10^{15}$~m$^{-2}$. 
We can see in Fig. 2, that a SdH dip appears as $B \sim B_{c}$, so the observed I-QH transition at $B_{c}$ is a direct one \cite{song, lee, huckestein}. In Fig. 2, magneto-oscillations cannot be observed at low $B$ until we increase the magnetic field to about $B=B_{c}$. Since such oscillations are due to Landau quantisation, the 2DES provides us an opportunity to probe the direct I-QH transition which occurs as Landau quantisation can just be identified. In addition, we can see that $\rho_{xx}=3.4$ k$\Omega \approx \rho_{xy}=3.1$ k$\Omega = \frac{B}{ne}$ at $B_{c}$ at $T=4$~K although the Hall slope is weakly $T$-dependent. So the observed transition occurs as 
$\rho_{xx}/\rho_{xy} \approx 1$, which seems to be consistent with Huckestein's argument.   
The low-field oscillations are expected to follow Eq. (3), the SdH formula. To analyze the mobility from Eq. (3), we 
note that ln$(\Delta \rho_{xx}/(\chi/$sinh$\chi))=$ const - $\pi/(\mu B)$.  We can see from the inset to Fig. 3 that 
the data of ln$(\Delta\rho_{xx}/(\chi/$sinh$\chi))-1/B$ at different temperatures collapse well into a single straight line when we take $m^{\ast}=0.067 m_{0}$ as the expected value in a GaAs 2DES. From the slope of ln$(\Delta\rho_{xx}/(\chi/$sinh$\chi))-1/B$, the quantum mobility $\mu =0.13$~m$^{2}$/Vs. Therefore, we can obtain the product $\mu B=0.46$ at the transition point $B=B_{c}$. Such a product deviates much from 1, and thus our result is inconsistent with Huckestein's argument although the direct I-QH transition occurs just as the magneto-oscillations due to Landau quantisation can be observed under Eq. (2). 

  It is known that Landau quantisation can result in magneto-oscillations as the product $\mu B<1$ \cite{tyhuangcm}. Therefore, the appearance of magneto-oscillations near $B_{c}$ does not indicate that the transition occurs just as Eq. (1) is valid. While numerical studies show that such transitions can occur just as $\mu B \approx 1$ in a non-interacting 2DES, Landau quantisation can induce magneto-oscillations at $\mu B<1$ where such a 2DES is an insulator \cite{nita1}. The coexistence of magneto-oscillations and insulating behaviour can be explained by the percolation theory \cite{shimshoni, trugman}. We note that Huckestein considered only a single mobility based on the Drude model, but another mobility  $\mu ^{\prime}$ has been introduced in Refs.\cite{li, cho, minkov, arapov}. The mobility $\mu$ corresponds to the quantum mobility while $\mu ^{\prime}$ can be related to the transport mobility although renormalization effects may be important \cite{minkov}. The direct I-QH transition should occur as $\mu ^{\prime} B=1$, and thus we can obtain $\mu ^{\prime} =1/B_{c}=0.29$~m$^{2}$/Vs $\approx 2.2 \mu$. Therefore, different mobilities should still be taken into account even as Landau quantisation can be identified just near $B_{c}$ with increasing $B$. 
   
   In Huckestein's argument, the direct I-QH transition separates the weak-localisation regime from the QH liquid due to the strong localisation under Landau quantisation. At low $B$, however, either Landau quantisation or the quantum Hall effect can be irrelevant to the strong localisation effect \cite{coleridge, hang, jhchen, coleridge2, tobias1, tobias2}. The onset of magneto-oscillations following Eq. (3) near the transition field $B_{c}$, in fact, does not indicate the importance of the strong localisation to the direct I-QH transition because Eq. (3) can hold without any localisation effect \cite{coleridge,Fogler}. Huckestein's argument is valid only if the onsets of both the strong localisation and Landau quantisation are at $\mu B \approx \mu ^{\prime} B \approx \rho_{xy}/\rho_{xx} \approx 1$. Our study shows that the direct I-QH transition does not always indicate the onset of strong localisation even when Landau quantisation can be identified just near the transition point with increasing $B$.             
  
    In Refs. \cite{li, cho, minkov}, quantum correction based on the e-e interaction is taken into account to explain why direct I-QH transition occurs at $\mu ^{\prime} B \approx 1$. We note that the e-e interaction effect can modify the 2D density of states near the Fermi level, giving rise to a logarithmic $T$-dependent Hall slope of a 2DES \cite{simmons}.
As shown in Fig. 3, the Hall slope is logarithmic $T$-dependent at $T=0.5-4$~K in our system. Since the carrier density determined from the oscillations in $\rho_{xx}$ remains constant over the same temperature range, the observed logarithmic $T$-dependent Hall slope can only be ascribed to e-e interaction effect within our system. The parabolic negative magneto-resistance, however, is not apparent at  $\mu B<1$ in Fig. 2 although it is also expected under the e-e corrections \cite{li}. In addition, we note that the magneto-oscillations are absent at $B_{c}$ in Ref. \cite{li} and Ref. \cite{cho}. while they appear near the transition point in our study and in Refs. \cite{song, lee}. In different 2D systems, therefore, it is possible that the dominant effects and/or parameters are not the same at low fields \cite{li, cho, gusev}. We can see from Fig. 3 that the Hall slope under a current $I=40$~nA somewhat deviates from the expected logarithmic $T$ dependence at the lowest temperature. To understand the mechanism for the deviation, we note that $\rho_{xx}$ at $B=0$ is $I$-dependent with increasing the current, as shown in the inset to Fig. 3. Here $\rho_{xx}(B=0)$ represents the value of $\rho_{xx}$ at zero magnetic field. The $I$-dependence indicates the existence of the current heating, under which the electron temperature $T_{e}$ is higher than the lattice temperature $T$ \cite{wennberg}. Therefore, effects due to electron-phonon interaction could be important in our study for electrons to transfer the extra energy to the lattice, which can induce the deviation of the Hall slope at low $T$. We note that the zero-field resistivity can be used as a self thermometer to determine the electron temperature $T_{e}$ \cite{wennberg}. It is expected that the Hall slope of a 2DES can also be used as a thermometer \cite{simmons}. As shown in Fig. 4 and its inset, both the zero-field resistivity and the Hall slope show that $T_{e} \propto I^{\alpha}$ with the exponent $\alpha  \approx 0.5$, which is expected under the electron-phonon interaction \cite{scherer}. Actually the low-field regime is unstable in the global phase diagram of the quantum Hall effect \cite{klz}, and more studies are necessary to clarify the dominant effects and/or parameters at low magnetic fields \cite{nita1, li, cho, minkov, arapov, kaminskii, martin, adamov, hang, jhchen, coleridge2, tobias1, tobias2}. 
    
    In our study, both $\mu$ and $\mu ^{\prime}$ remain the same after decreasing the driving current, which indicates that the current heating and/or electron-phonon interaction is irrelevant to the difference between these two mobilities. By decreasing the current to $I=12$~nA, as indicated by the red square in Fig. 3, the deviation on the logarithmic $T$-dependence of the Hall slope at low $T$ can be removed. In addition, we note that the direct I-QH transition at $\mu B=1$ can still be related to the e-e interaction effect when corrections to the negative magneto-resistance are taken into account. Moreover, the linear $T$-dependence of the inverse of the phase coherence time $\tau_{\phi}$ in Fig. 5 indicates the scattering due to the e-e interaction while the nonzero intercept shows the zero-temperature dephasing \cite{linandbird}. The phase coherence time $\tau_{\phi}$ is obtained by fitting our data to the low-field equation \cite{hikami}
  
\begin{eqnarray}
\Delta \sigma_{xx}(B)= \frac{-e^2}{\pi h}[\Psi(\frac{1}{2}+\frac{B_{0}}{B})-\Psi(\frac{1}{2}+\frac{B_{\phi}}{B})],
\end{eqnarray}
as shown in the inset to Fig. 5, where $\Psi$ is the digamma function and $B_{0}$ and $B_{\phi}$ correspond to transport and phase coherence rates, respectively \cite{simmons}. Therefore, the direct I-QH transition in our study could be dominated by the e-e interaction effect rather than the onset of Landau quantisation although different mechanisms should be introduced to understand the details.

  To further check Landau quantisation near direct I-QH transitions, we re-examine the data published in our previous report \cite{cfhuang}. In that report, we also investigated direct I-QH transitions, near which magneto-oscillations can be identified, at low magnetic fields in a gated 2DES. Magneto-oscillations can be observed as the filling factor  $\nu \sim 10$ in such a 2DES when the gate voltage $V_{g}=+ 0.15$ and 0 V, and we can apply Eq. (3) to analyze the quantum mobility after the appearance of I-QH transitions. Figure 6 (a) and (b) show the curves of ln$(\Delta \rho_{xx}/(\chi/$sinh$\chi))-1/B$ at these two gate voltages, and the slopes yield $\mu=0.53$ and 0.47 m$^{2}$/Vs at $V_{g}=+0.15$ and 0 V, respectively. On the other hand, the transition points yield $\mu ^{\prime}$ =1.9 and 1.7 m$^{2}$/Vs under these two gate voltages. The quantum mobility  $\mu$ is much smaller than the mobility $\mu ^{\prime}$ 
obtained from the transition point. Therefore, different mobilities should also be introduced to understand the direct I-QH transitions. 
  
    In conclusion, we investigate Landau quantisation and the direct I-QH transition in the two-dimensional electron system in an AlGaAs/GaAs heterostructure. Our study shows that such a transition does not occur as $\mu B=1$ even when Landau quantisation can be identified just near the transition point by the appearance of magneto-oscillations as  $\rho_{xy}/\rho_{xx} \approx 1$. Therefore, our study supports that different mobilities should be introduced for the direct I-QH transition and Landau quantisation. The temperature-dependences of the Hall slope and dephasing time indicate the importance of the effects of the e-e interaction to the direct I-QH transition although different mechanisms should be considered for the details of such a transition. The appearance of Landau quantisation or direct I-QH transition, in fact, does not always correspond to the onset of the strong localisation effect giving rise to quantum Hall liquids. 

   This work was funded by the NSC, Taiwan. We would like to thank Yu-Ru Li, Po-Tsun Lin, Yen Shung Tseng
and Chun-Kai Yang for their experimental help. K.Y.C. acknowledges a Taiwan/Japan exchange student grant provided by the NSC, Taiwan. We would like to thank Professor Efrat Shimshoni and Dr. Marian Nita for helpful discussions. 

Figure Captions

Figure 1 A Schematic diagram showing the sample structure

Figure 2 Longitudinal and Hall resistivity as a function of magnetic field ($B$) at various temperatures $T$. The dotted line indicates the transition point $B_{c}$. The inset shows ln$(\Delta \rho_{xx}/(\chi/$sinh$\chi))$ as a function of $1/B$ at $T= 0.7, 0.9, 1.1, 1.3, 1.5, 3$ and 
$4$~K, respectively.

Figure 3 Hall slope as a function of $T$. The black squares represents the result obtained at $I=40$~nA while the red square corresponds to the data obtained at $I=12$~nA. The straight line corresponds to the best linear fit at $T=0.5$ to $4$~K.
The inset shows the zero-field resistivity $\rho_{xx}$ as a function of current $I$ at various temperatures $T$. From top to bottom: $T=1.5, 1, 0.59,$ and $0.28$~K.

Figure 4 Logarithmic of electron temperature $T_{e}$ as a function of logarithmic of current $I$ determined from the zero-field resistivity $\rho_{xx}$ at different lattice temperature $T$. The best linear fit corresponds to $T_{e} \propto I^{\alpha}$ with $\alpha = 0.46$. The inset shows logarithmic of electron temperature $T_{e}$ as a function of logarithmic of current $I$ determined from the measured Hall slope $R_{H}$. The best linear fit corresponds to $T_{e} \propto I^{\alpha}$ with $\alpha = 0.53$.

Figure 5 The inverse of phase coherence time $1/\tau_{\phi}$ as a function of $T$. The inset shows 
$\Delta \sigma_{xx} = \sigma_{xx}(B)-\sigma_{xx}(B=0)$ as a function of $B$ (black curve). The red curve corresponds to the fit to Eq. (4). 

Figure 6 ln$(\Delta \rho_{xx}/(\chi/$sinh$\chi))$ as a function of $1/B$ at (a) $V_{g}= + 0.15$~V and (b) $V_{g}= 0$ at different temperatures $T$. 

\end{document}